\documentclass{emulateapj}

\begin{document}

\shortauthors{Luhman \& Potter}
\shorttitle{Testing Models with AB~Dor~C and the IMF}

\title{Testing Theoretical Evolutionary Models with AB~Dor~C and the 
Initial Mass Function\altaffilmark{1}}

\author{K. L. Luhman\altaffilmark{2} and D. Potter\altaffilmark{3}}

\altaffiltext{1}{
Based on observations made with ESO Telescopes at the Paranal Observatories
under program ID 60.A-9026. This publication makes use of data products
from the Two Micron All Sky Survey, which is a joint project
of the University of Massachusetts and the Infrared Processing
and Analysis Center/California Institute of Technology,
funded by the National Aeronautics and Space Administration
and the National Science Foundation.}

\altaffiltext{2}{Department of Astronomy and Astrophysics,
The Pennsylvania State University, University Park, PA 16802, USA; 
kluhman@astro.psu.edu.}

\altaffiltext{3}{Steward Observatory, The University of Arizona, Tucson, AZ
85721, USA; potter@as.arizona.edu.}

\begin{abstract}
In this paper, we assess the constraints on the evolutionary models of young
low-mass objects that are provided by the measurements of the companion
AB~Dor~C by Close and coworkers and by a new comparison of 
model-derived initial mass functions (IMFs) of star-forming regions to
the well-calibrated IMF of the solar neighborhood.
After performing an independent analysis of all of the imaging and 
spectroscopic data for AB~Dor~C that were obtained by Close,
we find that AB~Dor~C (which has no methane) is not detected at a significant 
level (signal-to-noise$\sim1.2$)
in the simultaneous differential images (SDI) when one narrow-band image 
is subtracted from another, but that it does appear in the individual SDI 
frames as well as the images at $J$, $H$, and $K_s$. 
Our broad band photometry for AB~Dor~C is consistent with that of Close.
However, the photometric uncertainties that we measure 
are larger than those derived by Close;
our uncertainties are consistent with those measured in other
studies using the same adaptive optics system.
Using the age of $\tau=75$-150~Myr recently estimated for AB~Dor by
Luhman, Stauffer, and Mamajek, the luminosity predicted by the models 
of Chabrier and Baraffe is consistent with the value that we estimate 
from the photometry for AB~Dor~C. 
We measure a spectral type of M6$\pm1$ from the $K$-band spectrum of AB~Dor~C,
which is earlier than the value of M8$\pm1$ reported by Close and is 
consistent with the model predictions when a dwarf temperature scale is adopted.
In a test of these evolutionary models at much younger ages, we show that the
low-mass IMFs that they produce for star-forming regions are similar to the 
IMF of the solar neighborhood.
If the masses of the low-mass stars and brown dwarfs in these IMFs of
star-forming regions were underestimated by a factor of two as suggested by
Close, then the IMF characterizing the current generation of Galactic 
star formation would have to be radically different from the IMF of the 
solar neighborhood.
\end{abstract}


\section{Introduction}
\label{sec:intro}

For decades, theoretical evolutionary models have been an essential tool
for interpreting the positions of stellar populations on the
Hertzsprung-Russell diagram.
In particular, studies of star and planet formation have relied
on these models to provide estimates of masses and ages of members of
star-forming regions ($\tau<10$~Myr). Because the uncertainties in the
models are largest at such early stages \citep{bar02}, observational tests
of the masses and ages inferred from the models have been crucial
\citep{whi99,luh99,sim00,pal01,bar02,luh03b,hw04,sta04}.
However, for young objects with masses below 0.3~$M_\odot$,
fewer tests of the mass estimates are available \citep{moh04,rei05}.
Recently, \citet{clo05} partially addressed this problem 
by measuring a dynamical mass for a low-mass companion, AB~Dor~C, 
that is much older than star-forming regions but still
above the main sequence.
This object was first detected indirectly by \citet{gui97} through astrometry 
of AB~Dor, which is a well-studied nearby young K-type star ($d=14.9$~pc). 
\citet{clo05} resolved AB~Dor~C from the primary with adaptive optics (AO)
images and used a combination of these data and the astrometry from 
\citet{gui97} to arrive at a mass of $M=0.09\pm0.005$~$M_\odot$.
After adopting an age of 50~Myr for the AB~Dor system, they found that 
the theoretical evolutionary models of \citet{cha00} overestimated the 
near-infrared (IR) fluxes of AB~Dor~C by roughly one magnitude. As a result,
\citet{clo05} concluded that these models underestimate the masses of 
young low-mass objects by a factor of two and that many of the objects 
previously identified as brown dwarfs in star-forming regions and open 
clusters are instead low-mass stars.

The observations of AB~Dor~C by \cite{clo05} (hereafter C05)
potentially provide
an important new test of the theoretical evolutionary models of 
low-mass objects. However, the results of this test hinge on reliable 
measurements of the mass, age, luminosity, and temperature of AB~Dor~C.
\citet{luh05b} recently reexamined the age constraints of the AB~Dor system, 
concluding that it has the same age as the Pleiades open cluster 
($\tau=100$-125~Myr), which removed the discrepancy between the luminosity 
reported by C05 and the value predicted by the models of 
\citet{cha00}.  In this paper, we reexamine other stellar properties of
AB~Dor~C, namely its luminosity and temperature, and describe the
constraints these data provide for evolutionary models (\S~\ref{sec:abdor}).
In addition, we present a second test of the models at low masses, but now at
much younger ages ($\tau\sim1$~Myr),
in which we compare the low-mass initial mass function (IMF) for nearby 
star-forming regions as derived with the models of \citet{bar98} and
\citet{cha00} to the relatively accurate and well-calibrated IMF of
the solar neighborhood under the assumption that the true IMFs of these 
populations are the same (\S~\ref{sec:imf}).

\section{AB~Dor C}
\label{sec:abdor}

\subsection{Photometry}

Given the large flux ratio and small angular separation of AB~Dor~A and C,
a robust photometric measurement of the latter with AO images requires careful 
consideration of the possible errors inherent to data of this kind. For these 
reasons, we have analyzed the data published in C05, which are now available 
to the public at the data archive of the European Southern Observatory (ESO).  

\subsubsection{SDI Images}
\label{sec:sdi}

The data with which C05 reported the first direct detection of AB~Dor~C
were obtained during
the commissioning of the simultaneous differential imaging mode 
\citep[SDI,][]{len04,bil04} of the High-resolution Near-IR Camera 
\citep[CONICA,][]{len98}, 
which was used in conjunction with the Nasmyth Adaptive Optics System 
\citep[NAOS,][]{rou00} on ESO's Very Large Telescope (VLT).
The SDI technique, first illustrated by \citet{rac99} and further 
developed by \citet{marois00,marois03,marois05},
uses the relative spectral contrast between a primary and a companion 
to suppress the contrast-limiting speckle noise. It has been 
shown that significant improvements in speckle suppression are gained 
through the SDI technique. However, non-common path errors, differential 
chromatic aberrations, alignment-sensitive ghost images, flat-field 
miscalibration, and detector read noise can significantly exceed
the photon noise. Although these data are not directly relevant to 
testing the theoretical evolutionary models, we examine them in this section
because of the importance of testing the abilities of new observational 
techniques like SDI.

\begin{figure}
\plotone{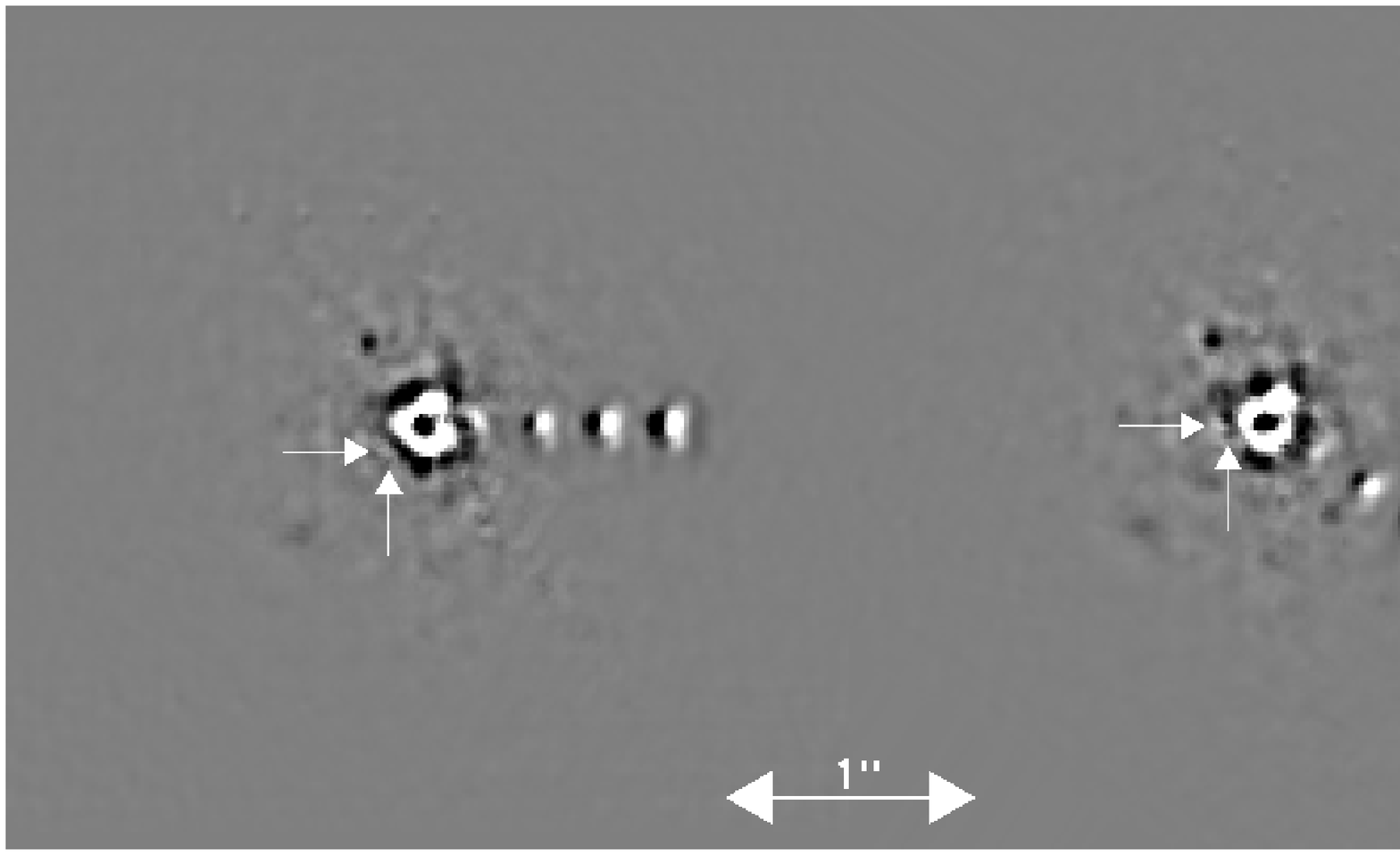}
\caption{
The differences of SDI images of AB~Dor~A at 1.625 and 1.575~\micron\ for 
the two rotator angles separated by $33\arcdeg$ produced by reprocessing the
data from \citet{clo05}. Artificial companions at the contrast 
level reported by \citet{clo05} for AB~Dor~C
have been inserted in the original images at four radii from AB~Dor
(0$\farcs$16, 0$\farcs$4, 0$\farcs$6, and 0$\farcs$8). 
The signature of a companion produced by this differential imaging 
technique is a difference between radially shifted PSFs.
The arrows indicate the location of AB~Dor~C reported by \citet{clo05}.
Neither AB~Dor~C nor the artificial companions with the same radius 
from AB~Dor~A are reliably detected.
North is up and East is left in the left image.
}
\label{fig:sdi}
\end{figure}

The SDI observations from C05 
were obtained on the night of 2004 February 1 and consisted of a set of images
at five dither positions followed by a second set of dithered images with the 
instrument rotator offset by $33\arcdeg$.
At a given position, four SDI images were obtained through narrow-band 
filters, one at 1.575~\micron, one at 1.600~\micron, and two at 1.625~\micron. 
After removing bad pixels and dividing by flat field images, we created images 
that were spatially scaled by the ratio of central filter wavelengths 
to make the diffraction and speckle pattern for all images coincident 
in image coordinates. The images were optimally 
aligned to subpixel accuracies using the flux conserving interpolation 
routine {\it interpolate} within IDL. After the subpixel alignment, 
the individual images were optimally flux scaled to minimize the residual 
counts in the non-saturated regions of the PSF halo in the subtracted image. 
The images were then spatially filtered as in \citet{mas05}.
The six different non-redundant 
subtraction combinations possible between the four images were created 
using these optimal subtraction routines for each of the 10 dither positions.
Because the plate scale is independent of the wavelength in the SDI mode of 
CONICA, when the images are spatially de-magnified to match the diffraction 
pattern of the PSFs with different wavelengths, the positions of a 
companion in different filters becomes radially misaligned to a degree 
proportional to the demagnification factor and the distance from the center
of the primary PSF.  As a result, the signature of a real (non-methane)
companion using the SDI technique is 
not just a fainter version of the primary PSF but rather a subtraction 
between two radially misaligned PSFs, as illustrated in Figure~\ref{fig:sdi}.
This effect essentially improves the sensitivity to the detection of 
companions by producing a more unique and detectable signal 
even when the companion has no intrinsic spectral contrast between the 
narrow-band filters. 

At the position of AB~Dor~C reported by C05, we find a point source appearing 
in all four of the unsubtracted images of the SDI data at a
signal-to-noise of $\sim3$.
However, this feature overlaps with the diffraction spider arm, 
making the differential imaging crucial in confirming it is as a real
companion with these data alone. Indeed, after the optimal alignment and 
subtraction routine was performed through the dataset,
we found no obvious signature of a companion at this 
location in any iteration of the optimally subtracted images. 
Although not done by C05, we performed an additional step to 
remove possible fixed noise from the reduced differential images by 
subtracting the reduced data sets taken at the different instrument 
rotator angles. This would produce a negative version of the companion's 
signal at the same radius as the companion, but separated in position 
angle equal to the instrument rotator angle change (33$\arcdeg$). 
From this analysis, we again found no obvious signature of a companion.
A similar type of reduction was performed by \citet{nie05}, resulting
in the images in the first panel of their Figure 4. In contrast to their
claims, we see no significant detection of a companion in those data.
To quantify the detection limits, a noise analysis was performed using the 
five dithered images for each rotation angle for each of the six subtraction 
combinations. 
Because the central five pixels surrounding AB~Dor~A were saturated,
a PSF for use with artificial companions could not be measured from these data.
However, prior to that sequence of images, C05 obtained a single short
exposure in which AB~Dor~A was not saturated. We adopt this PSF of AB~Dor~A
for artificial companions in the following analysis.
This use of a PSF from a short exposure tends to overestimate the Strehl ratio 
of the longer exposure, which results in a detection limit that is overly 
optimistic. Based on the spectra of late-M objects at a 
variety of ages \citep[e.g.,][]{cus05},
the maximum flux ratio expected for AB~Dor~C between the 
1.625 and 1.575~\micron\ narrow-band filters is $F(1.625)/F(1.575)\sim1.1$.
Therefore, we simulated AB~Dor~C with an artificial companion that was
125 and 112 times fainter than AB~Dor~A for the images at 1.625 and 
1.575~\micron, respectively. 

By combining the signal from the artificial 
companion with the standard deviation of the total flux in bins of 
linear size similar to the diffraction limit ($\sim4$~pixels) around a ring 
at the radius of AB~Dor~C reported by C05 ($\sim9$~pixels), we arrived
at an expected signal-to-noise ratio of 1.2 for AB~Dor~C.
We checked the validity of this measurement by inserting the artificial 
companion into the raw data at the separation of 0$\farcs$16 reported by
C05, repeating the reduction procedures, and visually
inspecting the resulting images. For comparison, we also included 
companions at larger separations of 0$\farcs$4, 0$\farcs$64, 
and 0$\farcs$88. The differenced SDI images containing these companions
are shown in Figure~\ref{fig:sdi}.
Although a companion of this contrast would easily be 
detected at separations beyond $0\farcs3$, it is not reliably detected 
at the radius of $0\farcs16$ reported by C05 for AB~Dor~C. 
Thus, we cannot reproduce the putative detection of a companion shown 
in Figure 1b from C05. 
Their image also does not exhibit the expected plus/minus asymmetry 
expected for SDI data of this kind.
Nor is that source evident in the difference between two instrument 
rotator angles in the first panel of Figure~4 from \citet{nie05}. However, 
\citet{nie05} does show a significant detection of AB~Dor~C in newer SDI data 
obtained after C05 when the companion had moved to a larger separation, which 
is consistent with the detection limits that we have measured.

\subsubsection{Broad-band Images}
\label{sec:broad}

In addition to the SDI data, C05 obtained near-IR broad-band images of
AB~Dor with CONICA and NAOS (NACO). 
For AB~Dor~C, C05 reported two~$\sigma$ errors of 
+0.19 and -0.24 at $J$, +0.13 and -0.15 at $H$, +0.12 and -0.15 at $K_s$.
These errors correspond to signal-to-noise ratios of 
10, 16, and 11 in the three bands, respectively,
which are substantially better than achieved in previous companion searches
with NACO \citep{bil04,mas05}.
To investigate this apparent discrepancy, in this section we present a
detailed, explicit analysis of the photometric errors for AB~Dor~C in the
data obtained by C05, including the possible systematic errors inherent to 
high contrast photometric measurements with AO observations. We also consider
the unpublished images in a narrow-band filter at 3.74~\micron\ 
that were obtained of AB~Dor with NACO.

\begin{figure}
\plotone{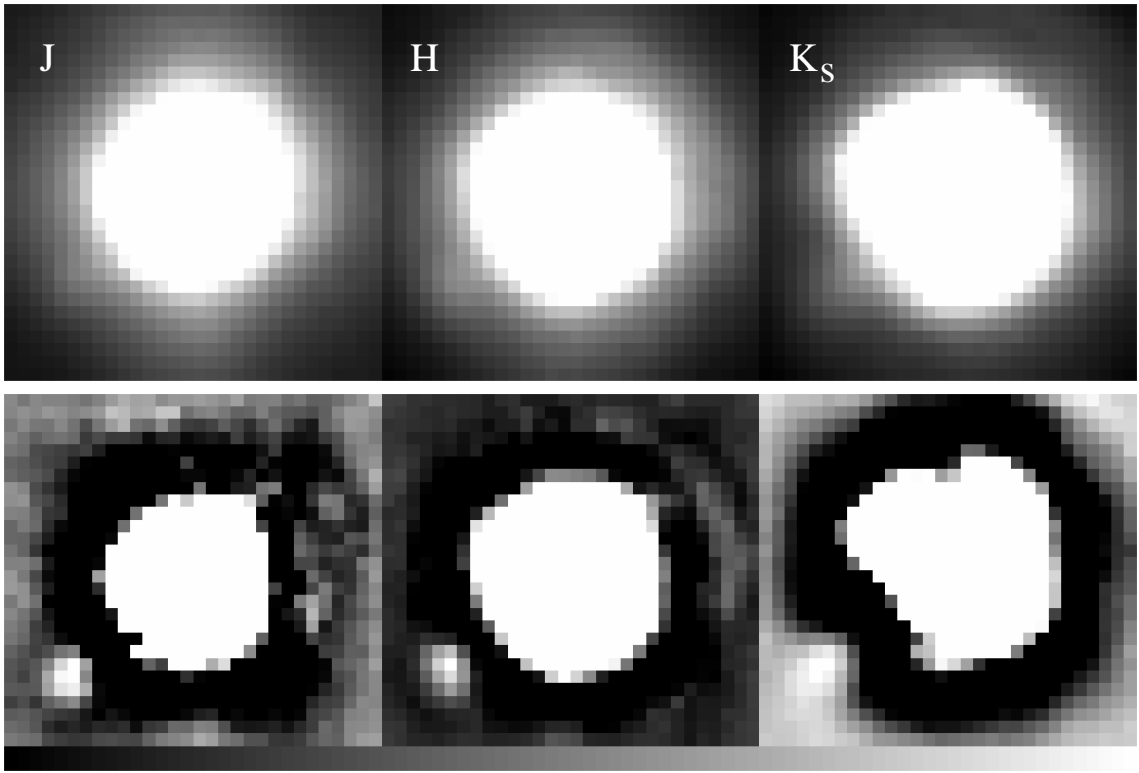}
\caption{
{\it Top:}
Near-IR images of AB~Dor~A ($0\farcs4\times0\farcs4$) 
produced by reprocessing the data from \citet{clo05}.
{\it Bottom:}
After applying a spatial filter to these images, AB~Dor~C is more 
noticeable. These filtered images are displayed linearly from the background
flux at the radius of AB~Dor~C to the maximum flux of this companion, 
as represented by the grayscale bar.
East is left and north is up in these images.
}
\label{fig:jhk}
\end{figure}

On the night of 2004 February 4, C05 obtained 15 images with exposure times 
of 10~s among 5 dither positions for each of the $J$, $H$, and $K_s$ filters
and three images with exposure times of 3.45~s (10 coadds of 0.345~sec).
A similar set of exposures was obtained through the NB3.74 filter, but
with exposure times of 3.6~s (10 coadds of 0.36~sec).
After flat fielding, dark subtraction, and removal of bad pixels, we 
constructed a sky frame in a given filter from a median combination of 
unaligned images and subtracted it from each image in that filter. 
We aligned the resulting images to subpixel accuracy using the same 
IDL interpolation routine as in the SDI data reduction.
In the long exposures, pixels
within a radius of $\sim0\farcs07$ from AB~Dor~A were saturated. 
These images were aligned using the unsaturated wings of AB~Dor~A. 
We also generated a set of images that were spatially filtered by 
subtracting from each image the same image convolved with a Gaussian with 
a width roughly equal to the FWHM for each wavelength \citep{mas05}.
In the resulting images, the low frequency PSF variations are suppressed.
The final nonfiltered and filtered images for $J$, $H$, and $K_s$ are
shown in Figure~\ref{fig:jhk}.

The companion AB~Dor~C is apparent in the individual reduced 10~s images
at $J$, $H$, and $K_s$, but not NB3.74.
A detection at NB3.74 is not expected given the total exposure time in that
filter. Although other fixed speckles and diffraction artifacts are present in
the halo of AB~Dor~A at comparable brightness, AB~Dor~C is the only 
"speckle" that remains at a constant position in the $J$, $H$, and $K_s$ images.
If it were a fixed diffraction speckle, the separation would be proportional to 
the central wavelength of the filter. The fact that it is in the same position 
in the three broad-band filters as well as the unsubtracted SDI images and is 
in all the different dither positions demonstrates the reality of the companion.

To measure photometry for AB~Dor~C from the $J$, $H$, and $K_s$ images, 
we used the shorter, unsaturated exposures of AB~Dor~A to construct a PSF
for each filter and calibrated it with photometry of AB~Dor~A from 2MASS.
We confirmed the accuracy of this calibration by extracting photometry for
AB~Dor~Ba+Bb, which appeared in some of the long exposures, and comparing
it to the 2MASS measurements. 
In each filter, the calibrated PSF was scaled and subtracted from the image at
the position of AB~Dor~C with a scale factor optimized such that the residual 
at the location of AB~Dor~C matched the light level of the adjoining regions.
The errors in these photometric measurements were computed from the standard 
deviation of the brightness at the position of AB~Dor~C in an area with a 
width equal to the PSF's FWHM through the 15 images.
In addition, we calculated the standard deviation of flux in FWHM-sized 
bins at all position angles at a constant radius from AB~Dor~A as done in 
the contrast sensitivity estimations of \citet{mas05}.  This analysis 
resulted in errors slightly larger than those calculated from the variance 
through a data set at the position of AB Dor C; specifically, the errors are 
12\%, 17\%, and 7\% larger using the method from \citet{mas05} for the 
$J$, $H$ and $K_s$ bands, respectively. 
This is likely due to the fact that the diffraction spikes 
from the secondary support create intrinsic variability around the PSF 
halo at a constant radius. For this reason, we use the lower error values 
calculated as the standard deviation of the flux through the data set at 
the position of AB~Dor~C.  
The photometry and errors produced by this analysis are presented in 
Table~\ref{tab:phot}. We also list the measurements from C05 for comparison.
The photometric errors for AB~Dor~C that we derive from C05's images
are significantly larger than those reported by C05. 
In comparison, the nonfiltered NACO data from \citet{mas05} 
exhibited an average sensitivity of 5~$\sigma$ at $\Delta m\sim4.5\pm0.5$
and $0\farcs16$, which corresponds to $\sim3$~$\sigma$ for an object at
the contrast of AB~Dor~C and thus is consistent with the photometric errors
we measure from the nonfiltered images.
Note that although the sensitivities from \citet{mas05} for filtered images
are better than those from the nonfiltered data, they represent only the 
significance of a detection rather than bona fide photometric errors 
because they do not include the systematic errors introduced by the filtering 
process.

In addition to the photometric errors listed in Table~\ref{tab:phot},
the noise associated with the variability of the reference PSF must also 
be taken into consideration.  The reference PSFs are from the short exposures 
of AB~Dor~A (0.345~s $\times$ 10 coadds $\times$ 3), which were
measured at a different time than the longer exposures that detected 
AB~Dor~C (10~s $\times$ 1 coadd $\times$ 15).
Because the total exposure time among the short exposures in a given band
was comparable to just one long exposure, we cannot directly measure the 
temporal variability of the PSFs used in the photometric analysis. 
However, the NB3.74 data, which were obtained on the same night as the 
long exposures, encompassed a longer time baseline. These data consisted
of a continuous series of 15 frames of 0.36~s $\times$ 10 coadds.
The combination of three consecutive images within this series has a similar
total clock time as one long exposure. Therefore, from the series of 15 images
we were able to construct 5 independent unsaturated PSFs that were measured
on the same timescale as the long exposures.
The variability of the NB3.74 PSF can then be translated into 
variabilities at other wavelengths using a relation between the Strehl 
ratio ($S$), the RMS wavefront error ($RMS$), and the observing wavelength 
$\lambda$: $S=exp(-(2\pi RMS/\lambda)^2)$. Defining the fractional error in 
the Strehl ratio to be $f=(S+\sigma_S)/S$, the fractional error 
relationship between two wavelengths then can be arrived at through 
algebraic manipulation: 
$f(\lambda_1)=f(\lambda_2)^{(\lambda_2^2/\lambda_1^2)}$. The standard 
deviation of the Strehl ratio of the 5 NB3.74 images was measured to be 
$f(\lambda_{NB3.74})=1.028$. To match the aperture used in the photometric 
analysis, we estimated the changes in the Strehl ratio using the total 
counts within the FWHM core of the PSF. The $f(J)$, $f(H)$, and $f(K_s)$ 
fractional Strehl ratio errors are found to be 1.286, 1.155, and 1.087 
respectively. These errors were propagated with those found from the halo noise
at the radius of AB~Dor~C and are presented in Table~\ref{tab:phot}.

\subsection{Spectral Type}
\label{sec:spt}

C05 used the spectroscopic mode of NACO to obtain near-IR spectra 
of AB~Dor~C on the night of 2005 February 4. 
The NACO system was configured to provide a slit width of $0\farcs086$, 
a spectral coverage of 2-2.5~\micron, and a resolution of 
$R\equiv\lambda/\Delta\lambda\sim1500$.
Eight dithered one-minute exposures were obtained with the slit aligned
along AB~Dor and the position of AB~Dor~C measured by C05,
and another set of eight exposures were taken with the slit rotated 
by $180\arcdeg$. The primary was saturated in these data. 
C05 and \citet{nie05} described some of the details of their processing
of these data.

\begin{figure}
\plotone{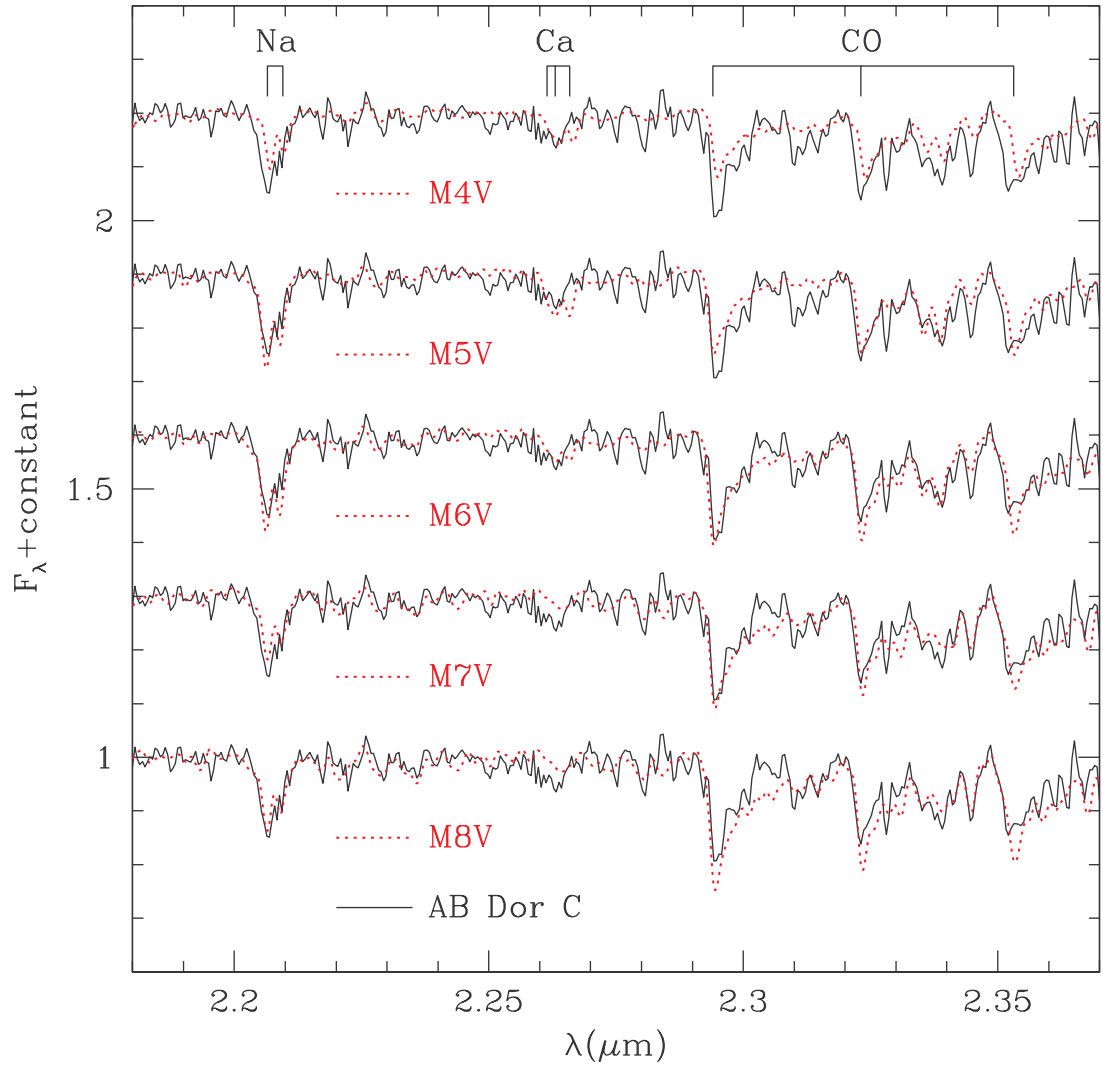}
\caption{
The spectrum of AB~Dor~C produced by reprocessing the data of \citet{clo05}.
By comparing this spectrum to data for field dwarf standards from 
\citet{cus05}, we measure a spectral type of M6$\pm1$ for AB~Dor~C. 
The standard spectra have been smoothed to the same resolution as the
data for AB~Dor~C ($R\sim1500$). All spectra have been divided by polynomial 
fits to their continua.
}
\label{fig:spec1}
\end{figure}

We have retrieved the NACO spectroscopic data for AB~Dor~C from the 
ESO archive and have performed an independent analysis, which we now describe.
After dividing by flat field images, we straightened the images to precisely
align the dispersion and spatial directions with the image coordinate axes. 
To detect AB~Dor~C within the wing of the PSF of AB~Dor, we
differenced spectral images at opposite instrument rotator angles.
To account for changes in the PSF, prior to this differencing we
measured the fluxes as a function 
of wavelength between radii of 8.5 and 12.5 pixels on the left and right
sides of AB~Dor for each of the eight images at each of the two rotator angles,
which we refer to as $F_\lambda (i,j,k)$ where $i=$left or right, 
$j=1$-8, and $k=0$ or 180 (degrees). We then computed
$r_\lambda(left)=F_\lambda(left,j1,0)/F_\lambda(left,j2,180)$ and 
$r_\lambda(right)=F_\lambda(right,j1,0)/F_\lambda(right,j2,180)$ 
for all combinations of $j1$ and $j2$. These ratios measured the change in the
PSF on each side of AB~Dor between a given pair of exposures. 
To achieve a reliable PSF subtraction, the PSF change should be the same on
each side of AB~Dor. Therefore, we considered only pair subtractions in 
which $r_\lambda(left)/r_\lambda(right)$ was approximately constant with 
wavelength. For each of these retained pairs, we fit a function to
$r_\lambda(left)+r_\lambda(right)$ that was linear with wavelength, 
multiplied the exposure at the second rotator angle ($k=180$) by this fit, 
and subtracted the resulting image from the image at the first angle ($k=0$). 
This process produced 9 good detections of AB~Dor~C out of the 16 that
were possible. We extracted these 9 spectra, combined them, and divided
by the spectrum of a telluric standard star, HD~34286 (G3V).
The resulting spectrum was multiplied by the solar spectrum to correct for 
the spectral slope and absorption features of this standard.
Because the 9 individual spectra of AB~Dor~C exhibited significant 
differences in their spectral slopes, the slope of the combined spectrum 
probably was not accurate.
We removed the slope of the spectrum by dividing by a polynomial fit to the 
continuum. The final spectrum of AB~Dor~C is shown in Figure~\ref{fig:spec1}.

For the spectral classification of AB~Dor~C, we use
the dwarf standards obtained with SpeX at the NASA Infrared
Telescope Facility by \citet{cus05}.
The resolution of the SpeX data is $R\sim2000$, which is higher than 
that of the NACO spectrum of AB~Dor~C ($R\sim1500$).
To enable a reliable comparison of the spectral features in these spectra,
we smoothed the SpeX data to the same resolution as the NACO data.
To do this, we compared the Ar lamp spectra obtained with the SpeX
and NACO spectra and identified a Gaussian function convolution for the SpeX
data that produced the same FWHMs as in the NACO spectra. We then applied
this Gaussian function to the spectra of the M dwarfs from \citet{cus05}.
The resulting SpeX data of Gl~213 (M4V), Gl~51 (M5V), Gl~406 (M6V), vB~8 (M7V),
and vB~10 (M8V) are plotted with the NACO spectrum of AB~Dor~C in 
Figure~\ref{fig:spec1}. 
The slopes of these standard spectra have been removed in the same manner
as for AB~Dor~C.

We now measure the spectral type of AB~Dor~C by comparing 
the strengths of the Na~I doublet, Ca~I triplet, and CO band heads 
in its spectrum to those exhibited by the dwarf standards in 
Figure~\ref{fig:spec1}.
The Na~I absorption in AB~Dor~C is weaker than that of M5V and M6V
and stronger than that of M4V, M7V, and M8V. Because this doublet is weaker
in pre-main-sequence objects than in dwarfs \citep{luh98a},
the strength observed for AB~Dor~C is consistent with types of M5-M6, 
but not M4, M7, or M8. In terms of both Ca~I and CO, AB~Dor~C is best matched 
by M6V. Compared to M4V and M5V, the lower surface gravity of AB~Dor~C 
could explain its stronger CO, but not the weaker Ca~I \citep{lr98}.
Meanwhile, the opposite is true for M7V and M8V. In particular, 
CO is significantly weaker in AB~Dor~C than in M8V, and the inconsistency 
is even larger when the gravity differences are considered.
Taken together, the strengths of Na~I, Ca~I, and CO indicate a spectral
type of M6$\pm1$ for AB~Dor~C.

Based on the spectrum of AB~Dor~C that they measured from the NACO data,
C05 and \citet{nie05} derived a spectral type of M8$\pm1$ for AB~Dor~C. 
We comment on a few aspects of their classification.
In agreement with our results in Figure~\ref{fig:spec1}, 
C05 found that AB~Dor~C exhibited stronger Na~I than vB~10 (M8V).
However, they attributed this difference to the lower surface gravity
of AB~Dor~C, even though Na~I transitions in general 
\citep{mar96,luh99,gor03,mc04} and this $K$-band doublet specifically
\citep{luh98a} are known to become weaker, not stronger, with
decreasing surface gravity. The stronger Na~I absorption in the
spectrum from C05 relative to M8V can only be explained 
by a spectral type that is earlier than M8V since the strength of this feature 
reaches a maximum at M6V and decreases with later types \citep{lr98,cus05}.
In addition to dwarfs, \citet{nie05} used young objects in Upper Scorpius
as standards. Compared to a spectrum of a M7 member of that association 
\citep{gor03}, AB~Dor~C exhibited stronger Na~I and CO, which \citet{nie05}
cited as evidence for a spectral type later than M7.
However, as already noted, stronger Na~I would imply a type earlier than
M7, not later. Meanwhile, the difference in CO is likely due to the fact that
the spectral resolution of the data for Upper Scorpius from \citet{gor03}
was much lower than the resolution for AB~Dor~C.

C05 also found good agreement between the broad spectral shapes 
(i.e., steam absorption) of their spectrum of AB~Dor~C and a spectrum of
vB~10 from \citet{wgm99}, which seemed to support
an M8 classification for the former. However, the validity of this comparison
is questionable for the following reasons. 
First, C05 corrected for telluric absorption in their data
by dividing by the spectrum of a solar-type star. They multiplied this result
by the solar spectrum to remove the spectral slope and absorption features
intrinsic to the telluric standard. In comparison, \citet{wgm99}
also divided their spectrum of vB~10 by a telluric standard (an A0 star),
but did not attempt to remove the intrinsic spectrum of that standard.
Thus, the procedures of C05 and \citet{wgm99} should produce
spectral slopes that differ systematically, which would preclude a meaningful
comparison of the steam bands in the resulting spectra of AB~Dor~C and vB~10.
In addition, \citet{nie05} stated that they (and presumably C05) were 
unable to ``preserve the continuum" of AB~Dor~C during their data reduction, 
which also would obviate the use of steam in the spectral classification.

It is useful to consider the spectral type of AB~Dor~C in the context of the
low-mass Pleiades members PPL15~A and B, which have well-constrained
masses and spectral types and have similar ages as AB~Dor~C.
According to the spectral type of M8$\pm1$ reported by C05,
AB~Dor~C should be cooler than PPL15~A and B \citep[M6 and M7;][]{bas99}. 
However, the dynamical mass of AB~Dor~C ($M=0.09\pm0.005$~$M_\odot$, C05)
is equal to or greater than the masses of PPL15~A and B, each of which have 
a firm upper mass limit of 0.1~$M_\odot$ based on the presence of
Li \citep{bas96,bas99}\footnote{This mass limit is based
on theoretical relationships between Li abundance, mass, and age, 
which have been shown to be robust \citep{bil97,bur04}.}
and are probably brown dwarfs based on their binary data \citep{bas99}.
Thus, these relative spectral types and masses of AB~Dor~C and PPL15 are 
incompatible. Meanwhile, the spectral type of M6$\pm$1 that we measure
for AB~Dor~C is perfectly consistent with PPL~15.

\subsection{Age}
\label{sec:age}

\begin{figure}
\plotone{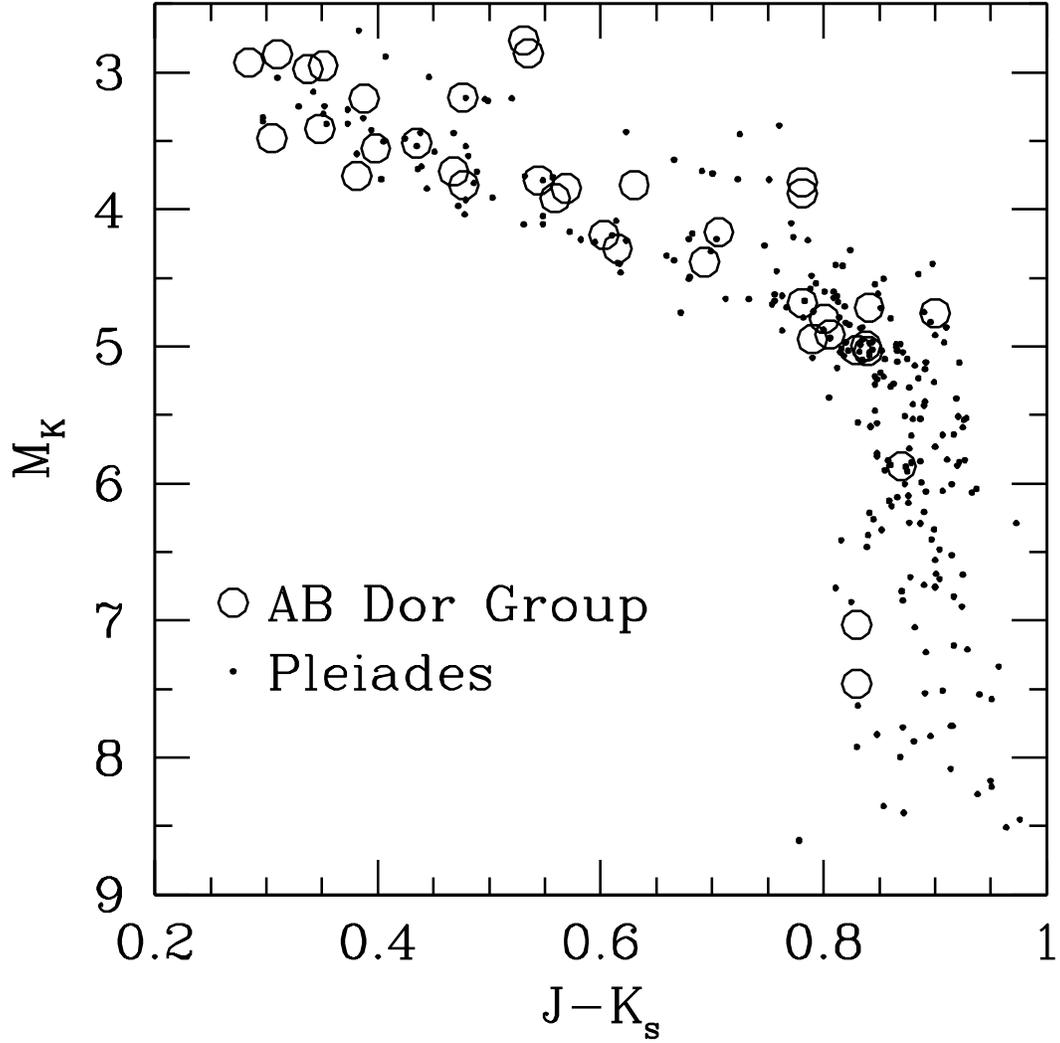}
\caption{
M$_K$ versus $J-K_s$ for members of the AB~Dor moving group 
\citep[{\it circles};][]{zuc04} 
and the Pleiades open cluster 
\citep[{\it points}, $\tau=100$-125~Myr,][]{meynet93,stauffer98a}.
The mean $M_K$ offset between the lower envelope of the Pleiades sequence
and the observed positions of stars in the Pleiades
and the AB~Dor moving group are the same to within $\sim0.05$~mag, indicating
that these two stellar populations have similar ages, just as found with
$M_K$ versus $V-K_s$ by \citet{luh05b}.
}
\label{fig:jk}
\end{figure}

\citet{luh05b} recently derived an age for AB~Dor from a color-magnitude 
diagram and the kinematics of the AB~Dor moving group.
In a diagram of $M_K$ versus $V-K_s$, they found that the AB~Dor group 
and the Pleiades were approximately coeval
\citep[$\tau=100$-125~Myr,][]{meynet93,stauffer98a}. This result was
supported by the kinematic analysis, which suggested that the two populations 
originated in the same large scale star-formation event.
\citet{nie05} briefly addressed the age of AB~Dor and the results of
\citet{luh05b}.  
First, they quoted \citet{luh05b} as finding that the AB~Dor group is brighter
by 0.1~mag than the Pleiades in $M_K$ versus $V-K_s$. However, that was not
the case. Instead, \citet{luh05b} reported an offset of 0-0.1~mag in a 
visual comparison of the histograms of the $M_K$ offsets between the observed 
positions of stars in each population
and a fit to the lower envelope of the Pleiades sequence.  In a more 
quantitative and definitive comparison, \citet{luh05b} found that the mean
offsets of the two populations were indistinguishable, indicating no 
detectable age difference.
\citet{nie05} then presented a comparison of the Pleiades and the AB~Dor group 
sequences in $M_K$ versus $J-K_s$ and claimed to find an offset of 0.15~mag
in $M_K$ between the sequences at $J-K_s>0.4$. However, a diagram of that kind
is a poor choice for measuring ages and comparing sequences because the 
dynamic range of $J-K_s$ is very small, nearly an order of magnitude lower than
that of $V-K_s$. In fact, the sequences of stellar populations are vertical in 
$M_K$ versus $J-K_s$ for low-mass stars later than K7, 
as illustrated in Figure~\ref{fig:jk}.
As a result, the stars that are normally most valuable for measuring ages
because of their large displacements above the main sequence are made 
useless by plotting them in $M_K$ versus $J-K_s$.
Finally, even if the AB~Dor group and the Pleiades are compared in 
$M_K$ versus $J-K_s$, as we have done in Figure~\ref{fig:jk}, their sequences
are still very similar. 
Following the procedure used for $M_K$ versus $V-K_s$ by \citet{luh05b},
we compared the mean offsets of the stars in the AB~Dor group and the
Pleiades from the lower envelope of the Pleiades. The difference between
these mean offsets ranges between -0.05 and 0.05 depending on the exact 
range of colors that is considered. Thus, the mean offsets for the AB~Dor
group and the Pleiades are indistinguishable in $M_K$ versus $J-K_s$, 
in agreement with the results from $M_K$ versus $V-K_s$,
and we find no basis in the work of \citet{nie05} for modifying
the age estimate of AB~Dor from \citet{luh05b}.

\subsection{Comparison to Model Predictions}
\label{sec:compare}

We now estimate the bolometric luminosity and effective temperature of 
AB Dor~C from our broad-band photometry and spectral classification and compare
them to the values predicted by the evolutionary models of \citet{cha00}.

\begin{figure}
\plotone{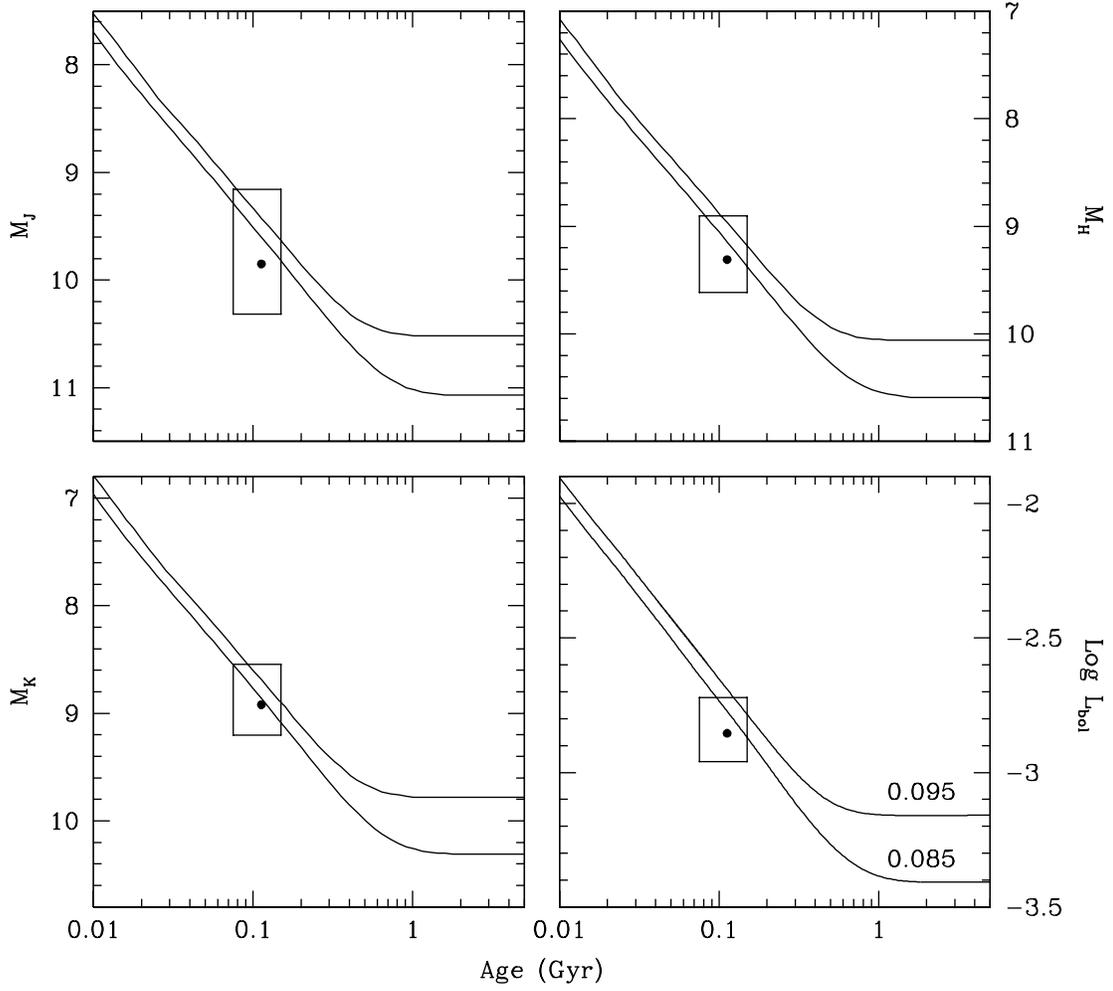}
\caption{
Comparison of our measurements of the near-IR magnitudes and bolometric
luminosity of AB~Dor~C to the 
values predicted by the evolutionary models of \citet{cha00} for masses
bracketing its dynamical mass of $0.09\pm0.005$~$M_\odot$ \citep{clo05}.
Using an age of 75-150~Myr for the AB~Dor system \citep{luh05b}, 
the predictions are consistent with the measurements.
}
\label{fig:lbol}
\end{figure}

We have computed the luminosity by combining our corrected $H$ magnitude 
in Table~\ref{tab:phot}, the distance of AB~Dor \citep{per97},
the $H$-band bolometric correction of M6 field dwarfs 
\citep[BC$_H=2.6$,][]{tin93,dahn02}, and an absolute bolometric magnitude
of 4.75 for the Sun, arriving at 
$L_{\rm bol}=0.0014^{+0.0005}_{-0.0003}$~$L_\odot$.
Because the $J-H$ and $H-K_s$ colors of AB~Dor~C are consistent with
those of M6 field dwarfs, a luminosity derived from $J$ or $K_s$ produces
a similar result. In Figure~\ref{fig:lbol}, we compare 
our measurements of $M_J$, $M_H$, $M_K$, and $L_{\rm bol}$ for AB~Dor~C to 
the values predicted by the theoretical evolutionary models of \citet{cha00}.
For AB~Dor~C, we adopt an age of 75-150~Myr \citep{luh05b}.
Using this age, the luminosity measured by C05 agrees well with the 
model predictions, as shown by \citet{luh05b}, while the agreement is slightly
worse using our new (smaller) luminosity estimate. However, the predicted 
luminosities are still well within the one $\sigma$ uncertainties of our 
measurement.  Similarly, each of the near-IR magnitudes that we measured for 
AB~Dor~C is consistent with the model values within the (large) photometric 
uncertainties. It is likely that deficiencies do exist at smaller levels
in the predicted parameters, particularly the IR magnitudes
\citep[e.g.,][]{cha00,leg01}.

\begin{figure}
\plotone{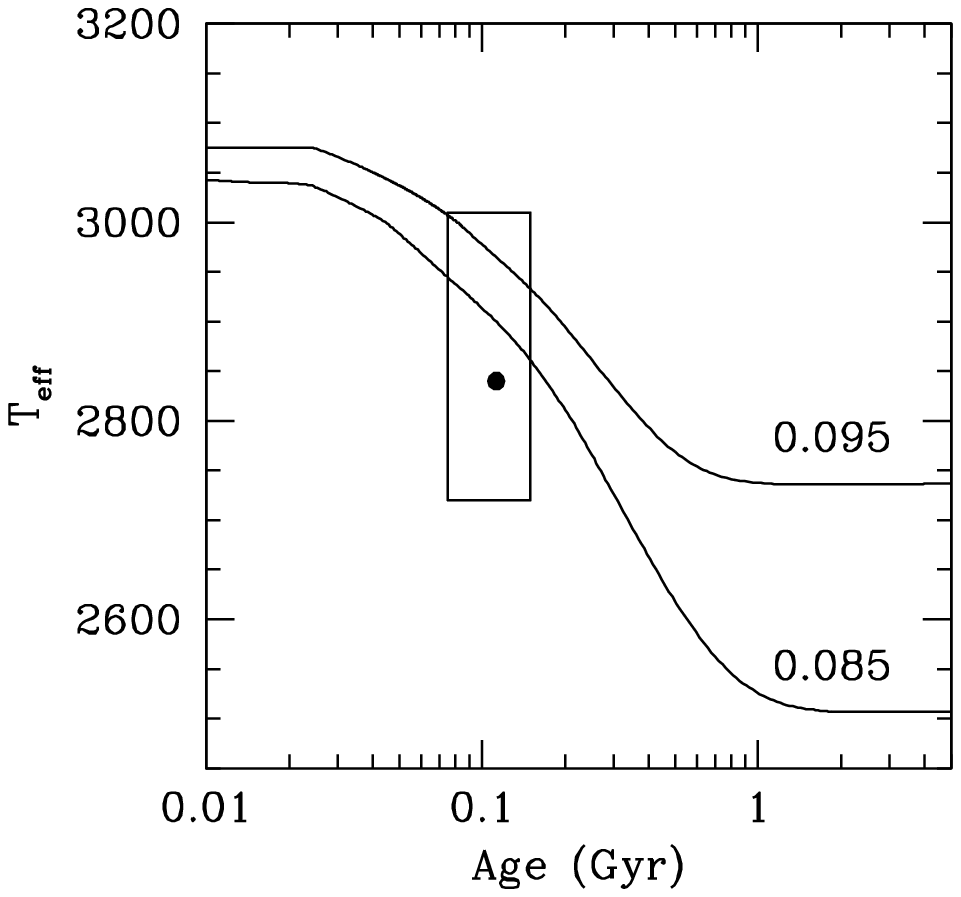}
\caption{
Comparison of our measurement of the spectral type of AB~Dor~C (M6$\pm$1) to
the temperatures predicted by the evolutionary models of \citet{cha00} for
masses bracketing its dynamical mass of $0.09\pm0.005$~$M_\odot$ \citep{clo05}.
Using the dwarf temperature scale compiled by \citet{luh99} and 
an age of 75-150~Myr for the AB~Dor system \citep{luh05b}, 
the predicted temperatures are consistent with the measured spectral type.
}
\label{fig:teff}
\end{figure}

To compare our spectral classification of AB~Dor~C to the model predictions,
we must adopt a conversion of spectral types to temperatures.
Using the dwarf temperature scale from \citet{luh99}, which was a fit to data
from \citet{leg96}, the spectral type of M6$\pm1$ for AB~Dor~C corresponds 
to $T_{\rm eff}=2840^{+170}_{-120}$~K. 
This estimate does not include the uncertainty in the 
M dwarf temperature scale, which is non-negligible.
An additional uncertainty is present for AB~Dor~C because a dwarf scale,
even if perfectly determined, may not apply to young objects \citep{luh99}. 
Nevertheless, even with underestimated errors, 
it is useful to compare our temperature estimate for AB~Dor~C to the model
predictions. As shown in Figure~\ref{fig:teff}, our spectral classification
of AB~Dor~C is consistent with the temperatures predicted by the models
of \citet{cha00} when the dwarf scale from \citet{luh99} is adopted.

\section{Initial Mass Functions of Star Forming Clusters and the Solar
Neighborhood}
\label{sec:imf}

The results of tests of evolutionary models at ages of $\sim100$~Myr like 
the one supplied by AB~Dor~C do not necessarily extend to much younger ages. 
The uncertainties in the evolutionary calculations become larger
at younger ages, and thus it would not be surprising if the models were robust
for relatively evolved objects like AB~Dor~C that are approaching the main
sequence but had significant errors at ages of a few million years
\citep{bar02}. Model tests that apply directly to the youngest ages are needed.
In this section, we present such a test of the masses estimated with the
evolutionary models of \citet{bar98} and \citet{cha00} for low-mass stars
and brown dwarfs in star-forming regions.

The Galactic disk appears to be populated predominantly by stars born 
in embedded clusters rather than in isolation 
\citep[][references therein]{lada03}.
Therefore, unless the IMFs of embedded clusters have changed significantly 
during the lifetime of the Galaxy, the current generation of clusters 
should have the same average IMF as the solar neighborhood.
Among the best-studied star-forming clusters, the IMFs of stars and 
brown dwarfs inferred from a given set of evolutionary models exhibit relatively
little variation \citep{luh00}. 
In logarithmic units where the Salpeter slope is 1.35,
the IMFs in most of these clusters rise from high masses down to a solar
mass, rise more slowly down to a maximum at 0.1-0.2~$M_\odot$, and then
decline into the substellar regime, as in 
the Orion Nebula Cluster \citep[$\tau\sim0.5$~Myr,][]{hil97,luh00b,hc00,mue02} 
and IC~348 \citep[$\tau\sim2$~Myr][]{her98,luh98b,luh03b,mue03}.
The only clear exception is the IMF in the Taurus star-forming 
region ($\tau\sim1$~Myr), which peaks near 0.8~$M_\odot$ \citep{bri02,luh04}, 
but quiescent, low-density regions like Taurus 
produce an insignificant number of stars compared to the giant
molecular clouds that contain most embedded clusters \citep{lada03}. 
Therefore, we can compare the model-derived IMFs of star-forming clusters 
to the IMF of the solar neighborhood to check the accuracy of mass 
estimates for young low-mass stars and brown dwarfs.

To represent the IMF of a typical embedded cluster, we select the IMF of 
IC~348 from \citet{luh03b}, which was derived with an H-R diagram of the 
cluster, the evolutionary models of \citet{bar98} and \citet{cha00}, and the 
temperature scale described in \citet{luh03b}. 
That IMF consists of primaries plus companions at projected separations 
greater than 300~AU, and thus more closely resembles an IMF of primaries
rather than a single star mass function. For the IMF of stars in the solar 
neighborhood, we adopt the measurement by \citet{rei02a} with a modification to 
exclude companions with projected separations less than 300~AU, making it 
suitable for comparison to IC~348. Similar measurements of the 
IMF of field stars have been presented by \citet{kro01} and \citet{cha01,cha03}.
Because the mass-luminosity relation is a function of age for brown dwarfs
at any age, and the ages of individual field brown dwarfs are unknown, 
a unique, well-sampled IMF of field brown dwarfs cannot be constructed. 
Therefore, we compare only the IMFs of low-mass stars 
between the solar neighborhood and IC~348. 
The available constraints on the substellar IMF for the field 
\citep{rei99,cha02,all05} have been compared 
to measurements in star-forming regions in previous work \citep{luh04}.

\begin{figure}
\epsscale{0.8}
\plotone{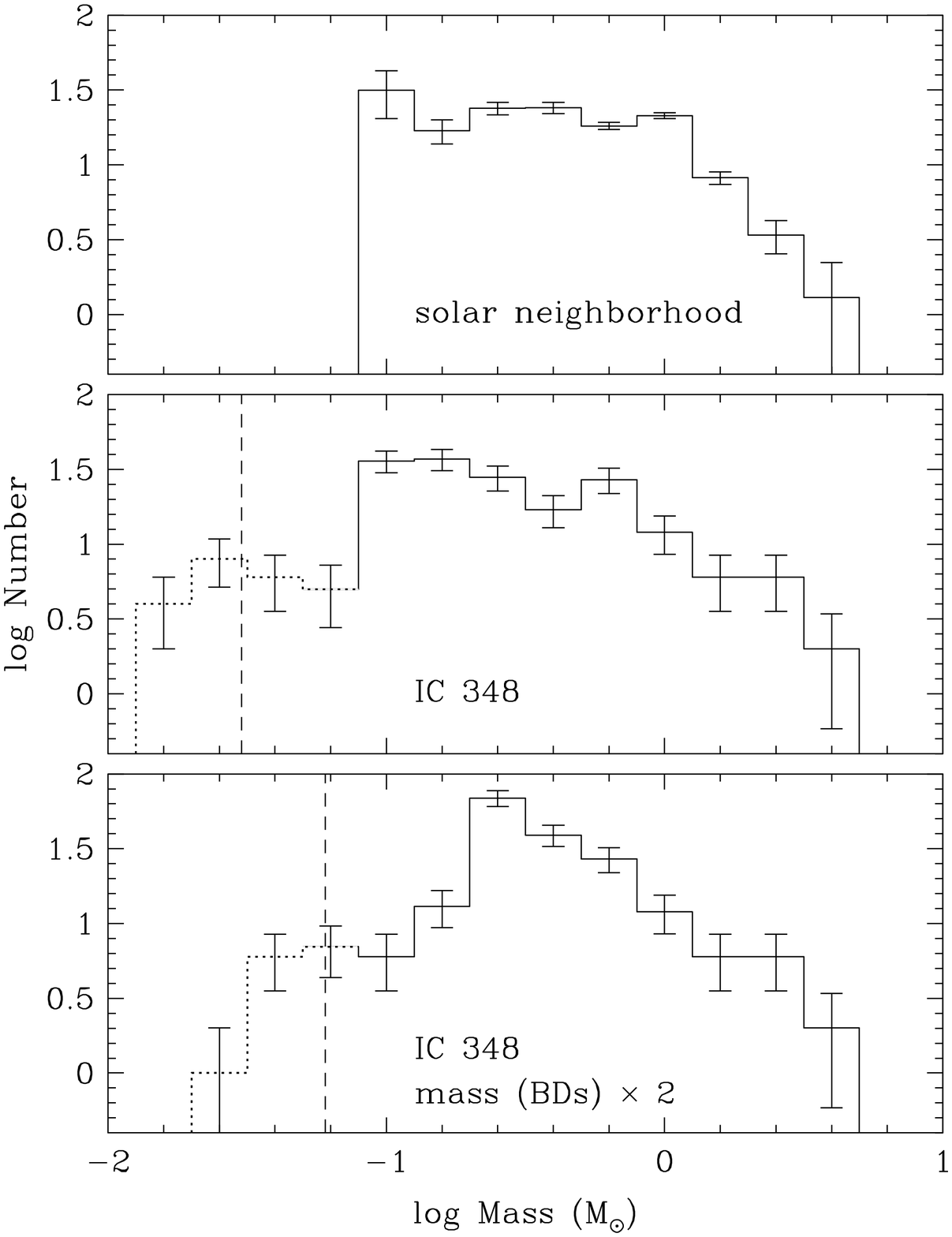}
\epsscale{1}
\caption{
Initial mass functions of stars in the solar neighborhood 
\citep[{\it top};][]{rei02a} and stars ({\it solid}) and brown dwarfs 
({\it dotted}) in IC~348, which is a typical Galactic star-forming cluster
\citep[{\it middle};][]{luh03b}.
The masses in the IMF for IC~348 were derived with the evolutionary
models of \citet{bar98} and \citet{cha00}. If the masses near and below the 
hydrogen burning limit in that IMF were significantly 
underestimated, then the corrected, true IMF for IC~348 ({\it bottom}) would 
be significantly different from that of the solar neighborhood.
The normalization of the IMF of the solar neighborhood is arbitrary.
The dashed line represents the completeness limit for IC~348.
In the units of this diagram, the Salpeter slope is 1.35.
}
\label{fig:imf}
\end{figure}

The IMFs for the solar neighborhood and the IC~348 star-forming cluster
are compared in Figure~\ref{fig:imf}. These IMFs are rather similar; both 
are roughly consistent with a Salpeter slope above a solar mass and are
slightly rising from a solar mass to 0.1~$M_\odot$.
Previous comparisons of the mass functions of young clusters and the solar 
neighborhood have arrived at the same conclusion \citep[e.g.,][]{cha03}.
This agreement supports the validity of the masses we have inferred from the
models of \citet{bar98} and \citet{cha00} for young low-mass objects. 
Meanwhile, C05 concluded that the masses of young low-mass members
of clusters 
have been underestimated by a factor of two with
the evolutionary models of \citet{cha00}. To investigate this possibility,
we have modified the IMF for IC~348 by doubling the mass estimates below 
0.08~$M_\odot$. Because our adopted temperature scale and evolutionary models 
produce accurate masses at $M\gtrsim0.5$~$M_\odot$ \citep{luh03b},
the factor by which the original masses are multiplied is selected to 
decrease linearly 
with log~M from two at 0.08~$M_\odot$ to unity at 0.5~$M_\odot$.
As shown in Figure~\ref{fig:imf}, the resulting IMF differs significantly 
from the IMF of the solar neighborhood. Most notably, the relative numbers
of stars at 0.25 and 0.1~$M_\odot$ differ by an order of magnitude between the
two IMFs. This exercise demonstrates that the masses derived for the low-mass
members of IC~348 cannot be underestimated by a factor of two if the IMF 
of Galactic star-forming clusters and the solar neighborhood are similar.
An alternative illustration of this result is the following. Our use of the 
evolutionary models indicates that young objects with optical spectral types
of M7 and later are brown dwarfs. Objects at these spectral types are 
relatively rare in star-forming regions \citep{bri02,luh03b,luh04}. 
In comparison, stars at masses of 0.1-0.2~$M_\odot$ are the most
abundant stars in the field.
Therefore, members of star-forming regions at types of M7 and later cannot 
be low-mass stars unless the IMFs of those regions and the solar 
neighborhood are radically different.

We have applied the above test only to the specific methods and models
used in estimating masses by \citet{luh03b}. The results of this test do
not apply to the synthetic near-IR magnitudes, other sets of evolutionary
models, or other temperature scales.

\section{Discussion}

C05 found that the evolutionary models of \citet{cha00}
overestimated the $J$ and $H$-band brightnesses of AB~Dor~C by one magnitude
and concluded that ``the young, cool objects hitherto thought to be 
substellar in mass are therefore about twice as massive, which means that 
the frequency of brown dwarfs and planetary mass objects in young stellar 
clusters has been overestimated" and that ``such errors will require
serious revision of the frequency for the lowest-mass objects...at young ages".
We find that this conclusion is oversimplified and unwarranted. First, the
error factor in the mass estimates for young low-mass objects is a strong 
function of the stellar parameters considered, the choice of evolutionary 
models, and other considerations, such as the adopted conversion between 
spectral types and temperatures. For instance, even in the analysis of C05,
the apparent error factor varied from 2 for $J$ and $H$ to 
only 1.3 for $K$. The conclusions of C05 included implications for measurements 
of substellar IMFs in star-forming regions, but they did not actually test 
the methods predominantly used to derive masses in those surveys. 
Instead, they considered mass estimates based on the synthetic
near-IR magnitudes, whereas bolometric luminosity 
is the most common parameter used in deriving masses for low-mass IMFs.
In fact, within the age and luminosity uncertainties quoted by C05,
the models of \citet{cha00} are formally consistent with the observations 
of AB~Dor~C, and so their sweeping conclusions regarding the 
validity of the models and of the measurements of substellar mass functions
were not justified by their own data.
Meanwhile, the new estimate of the age for AB~Dor~C from \citet{luh05b}
indicates reasonable agreement between its predicted luminosity and the values
measured by C05 and in this work (\S~\ref{sec:compare}).
In addition to the synthetic near-IR magnitudes, C05 compared
the effective temperature predicted for AB~Dor~C to the spectral type of
M8$\pm1$ that they measured from IR spectroscopy, and concluded that the models
overestimated its temperature.
However, we measure an earlier spectral type of M6$\pm1$ for AB~Dor~C from
C05's spectroscopic data, which is in agreement with the model predictions
if a dwarf temperature scale is adopted.
Purely on an observational basis, a spectral type earlier than the one
reported by C05 is expected given that AB~Dor~C has a mass greater than
or equal to that of PPL~15~A and B (M6 and M7).

We have also presented a simple test of the general possibility that the masses 
of young low-mass stars and brown dwarfs, particularly in star-forming regions 
($\tau\sim1$~Myr), have been significantly underestimated in previous 
studies using the models of \citet{cha00}.
This test consisted of a comparison of the model-derived IMFs of star-forming 
regions and the IMF of the solar neighborhood.  The agreement between
these IMFs supports the accuracy of the masses for young low-mass objects
when they are derived by combining measurements of spectral types and 
luminosities with the models of \citet{bar98} and \citet{cha00} and the 
temperature scale of \citet{luh03b}.
These mass estimates cannot have very large systematic errors ($\times2$) 
unless the IMF produced by the current generation of star formation in the 
Galactic disk is significantly different from the IMF of the solar neighborhood.
However, this test does not exclude smaller systematic errors.
Indeed, for the primary in a young spectroscopic binary in Upper Scorpius,
\citet{rei05} recently measured a dynamical mass of $>0.3$~$M_\odot$,
which is higher than the value of 0.2-0.3~$M_\odot$ 
derived with the same methods and models used in the above IMF measurements.
A more definitive test of those methods and models would be possible with 
a verification of the spectral type of that binary, as noted by \citet{rei05}. 
Dynamical masses at lower masses are also needed because 
the systematic errors could depend on mass \citep{moh04,rei05}.

\acknowledgments
We are grateful to Neill Reid for providing his data for the mass function of
the solar neighborhood. We thank Mike Cushing for providing the Ar lamp
spectra from SpeX. K. L. was supported by grant NAG5-11627 from the 
NASA Long-Term Space Astrophysics program.

\begin{deluxetable}{lccccccccc}
\tabletypesize{\scriptsize}
\tablecaption{Photometry of AB Dor C \label{tab:phot}}
\tablehead{
\colhead{Filter} &
\multicolumn{3}{c}{Close et al. 2005} &
\colhead{} &
\multicolumn{5}{c}{This work} \\
\cline{2-4} \cline{6-10}
\colhead{} & 
\colhead{S/N} &
\colhead{magnitude} &
\colhead{F(A)/F(C)} &
\colhead{} &
\colhead{S/N\tablenotemark{a}} &
\colhead{magnitude\tablenotemark{a}} &
\colhead{F(A)/F(C)\tablenotemark{a}} &
\colhead{magnitude\tablenotemark{b}} &
\colhead{F(A)/F(C)\tablenotemark{b}}}
\startdata
$J$ &     
10 & 10.76$^{+0.095}_{-0.12}$ & 150$\pm$15 & &
3.0  & 10.72$^{+0.31}_{-0.45}$ & 145$\pm$49 & 10.72$^{+0.40}_{-0.63}$ & 145$\pm$64 \\
$H$ &      
16 & 10.04$^{+0.065}_{-0.075}$ & 120$\pm$7.5 & &
4.2  & 10.18$^{+0.23}_{-0.30}$ & 136$\pm$33 & 10.18$^{+0.27}_{-0.37}$ & 136$\pm$39 \\
$K_s$ &   
11 & 9.45$^{+0.06}_{-0.075}$ & 80$\pm$7.5 & &
3.9 & 9.79$^{+0.25}_{-0.33}$ & 110$\pm$29 & 9.79$^{+0.27}_{-0.36}$ & 110$\pm$31 \\
\enddata
\tablenotetext{a}{In this measurement of AB~Dor~C in the long exposures, 
the short exposure of AB~Dor~A is used as the PSF.}
\tablenotetext{b}{Including errors from the variability of the PSF between
the short and long exposures (\S~\ref{sec:broad}).}
\end{deluxetable}
\clearpage

\end{document}